\documentclass[fleqn,usenatbib]{mnras}

\usepackage[T1]{fontenc}
\usepackage{ae,aecompl}

\usepackage[a4paper]{geometry}
\usepackage[utf8]{inputenc}
\usepackage{booktabs}
\usepackage{graphicx}
\usepackage{amsmath}
\usepackage{amssymb}
\usepackage[capitalise]{cleveref}
\usepackage{url}

\crefformat{equation}{Eq~(#2#1#3)}

\newcommand{\fR}{f\left(R\right)}

\title[Haloes in $\Lambda$CDM and $\fR$]{
The connection between halo concentrations and assembly histories: a probe of gravity?
}

\author[P. Ole\'{s}kiewicz et al.]{
	Piotr Ole\'{s}kiewicz$^{1}$\thanks{E-mail: piotr.oleskiewicz@durham.ac.uk},
	Carlton M. Baugh$^{1}$,
	Aaron Ludlow$^{2}$,
	\\
	$^{1}$Institute for Computational Cosmology, Department of Physics, Durham University, South Road, Durham DH1 3LE, UK \\
	$^{2}$International Centre for Radio Astronomy Research, University of Western Australia, 35 Stirling Highway, Crawley, \\
	Western Australia 6009, Australia
}

\date{Accepted XXX. Received YYY. in original form ZZZ}
\pubyear{2019}

\begin{document}
\label{firstpage}
\pagerange{\pageref{firstpage}--\pageref{lastpage}}
\maketitle

\begin{abstract}
	We use two high resolution N-body simulations, one assuming general
	relativity and the other the Hu-Sawicki form of $\fR$ gravity with $\vert
	\bar{f}_{R} \vert = 10^{-6}$, to investigate the concentration--formation
	time relation of dark matter haloes.  We stack haloes in logarithmically
	spaced mass bins to fit median density profiles and extract median
	formation times. At fixed mass, haloes in modified gravity
	are more concentrated than those in GR, especially at low masses and at low
	redshift, and do not follow the concentration--formation time relation seen
	in GR.  We assess the sensitivity of the relation to how concentration and
	formation time are defined, as well as to the segregation
	of the halo population by the amount of gravitational screening.  We find a
	clear difference between halo concentrations and assembly histories
	displayed in modified gravity and those in GR. Existing
	models for the mass--concentration--redshift relation that have gained
	success in cold and warm dark matter models require revision in $\fR$
	gravity.
\end{abstract}

\begin{keywords}
	dark matter -- galaxies: haloes -- methods: numerical
\end{keywords}

\section{Introduction}\label{sec:intro}

N-body simulations have driven astounding progress in improving our understanding of gravitational collapse and its role in the formation of cosmic structure and galaxy evolution. For example, simulations have demonstrated  that the mass distribution inside dark matter haloes follows an approximately universal form that can be specified by only two parameters \citep[][hereafter NFW collectively]{navarro1996,navarro1997}:
\begin{equation}\label{eqn:nfw}
	\frac{\rho\left(r\right)}{\rho_{\mathrm{crit}}}=
	\frac{\delta_{\mathrm{c}}}{\left(r/r_{-2}\right)\left(1+r/r_{-2}\right)^2},
\end{equation}
where $r_{-2}$ is a scale radius (at which the logarithmic slope of the density profile is equal to $-2$), and $\delta_{\mathrm{c}}$ is a characteristic overdensity. It is common to recast these into other forms, such as halo virial\footnote{We define the virial mass, $m_{200}=(800/3)\,\pi\,r_{200}^3\,\rho_{\rm crit}$, and corresponding virial radius, $r_{200}$, as that of a sphere (centred on the particle with the minimum potential energy) whose mean density is equal to 200 times the critical density, $200\times \rho_{\rm crit}$.} mass, $m_{200}$, and concentration, $c=r_{200}/r_{-2}$ (the ratio of the virial and scale radii). At fixed $m_{200}$, $\delta_{c}$ is given by
\begin{equation}\label{eqn:delta_c}
	\delta_{c} = \frac{200}{3}\frac{c^{3}}{\ln\left(1+c\right)-c/\l\left(1+c\right)},
\end{equation}
such that higher concentration implies higher characteristic density.

Simulations of structure growth in the cold dark matter model
(CDM) have also revealed a well-defined, redshift-dependent correlation between
these parameters: at fixed redshift concentrations decrease with increasing
mass, and at fixed mass decrease with increasing redshift \citep[see,
e.g.,][]{bullock2001,gao2008}. These trends betray an simpler relation
between the characteristic density of a halo and its formation time, $z_{\rm f}$: haloes that form early have, on average, higher $\delta_c$ than late-forming ones, reflecting the higher background density at that time  \citep[e.g.,][]{neto2007,ludlow2013}. This fact has been used to construct a number of empirical models for the  concentration-mass-redshift relation (hereafter $c(m,z)$, for short) that appeal to various definitions of formation time to predict characteristic densities, and hence concentrations \citep[e.g., NFW;][]{bullock2001,wechsler2002,zhao2003,maccio2008,zhao2009,ludlow2014a,correa2015,ludlow2016}.

Various models have met with varied success, plausibly due to
diverse definitions of collapse time \citep[see, e.g.,][for
details]{neto2007,ludlow2016}. Several studies define the formation time
of a halo as the point at which some fraction $F$ of its final virial mass had
first assembled, either into one {\em main} progenitor or accumulated over many
small progenitors. However, as first discussed in \citet{ludlow2013}, better
agreement with simulation results can be obtained by defining $z_{\rm f}$ in terms of
the halo's {\em characteristic} mass, $m_{-2}=m(<r_{-2})$, rather than
$m_{200}$ (we elaborate on this point in Section~\ref{sec:cmz}). This has
inspired a number of empirical models that successfully reproduce the $c(m,z)$
relation in both cold \citep{ludlow2014a,correa2015} and warm dark matter
cosmologies \citep{ludlow2016}

As a result, there exists an increasingly well-described relation between halo mass and concentration \citep{duffy2008,prada2012,angel2016,klypin2016,diemer2015,diemer2019}--the two parameters that are needed to specify the density profile of a relaxed dark matter halo--and how they evolve with time. Further, both analytic and empirical models have been shown to describe reasonably well the $c(m,z)$ relation for a variety of cosmological parameters and power spectra. Our objective here is to investigate whether the relation between concentration and formation time--upon which many of these models are based--is sensitive to the gravitational force law, as stark differences could be used to probe departures from general relativity.

Proposals for modifications to general relativity (GR) were originally motivated by trying to solve one of the biggest remaining problems with the concordance $\Lambda$CDM: the origin of the accelerated cosmic expansion.  $\Lambda$CDM achieves this by invoking a cosmological constant, $\Lambda$, but the required value is difficult to justify from a theoretical viewpoint \citep{carroll2004}. Many alternatives have been proposed to the standard $\Lambda$CDM model: the accelerated expansion could be driven by as-of-yet unknown physics in the dark sector \citep{zuntz2010} or by a modification to GR itself  \citep{koyama2016}.  Among the alternatives to GR, one of the most widely studied is $\fR$ gravity -- an umbrella term referring to modified gravity models which change the Ricci scalar in Einstein-Hilbert action \citep{buchdahl1970,clifton2012,joyce2015}. Current versions of the theory are fine-tuned to match the expansion history in $\Lambda$CDM, which removes some of the model's original appeal. Nevertheless, $\fR$ gravity remains a workable alternative to GR with interesting phenomenology.  While the parameter space of $\fR$ models is already tightly constrained by observations \citep{lombriser2014a}, there still exists a range of models which may display measurable differences from GR \citep[see, for example,][]{he2018,hernandez-aguayo2018}.

Our study uses the merger histories of dark matter haloes traced back to progenitors that are two orders of magnitude less massive than the final halo mass.  Hence, high resolution simulations are necessary (see \cref{tbl:nbody}). We therefore use the \textsc{Liminality} simulations of \cite{shi2015}, a suite of very high resolution dark-matter-only runs including examples of the \citet[HS]{hu2007} parametrisation of $\fR$ gravity.  Two simulations are compared: one of GR and another $\fR$ modified gravity model that is compatible with current observational constraints.

This paper is structured as follows.  The theoretical background is given in
\cref{sec:theory}: the $\fR$ model is discussed in \cref{sec:fr}, a description
of the $c(m,z)$ model of \citet{ludlow2016} in \cref{sec:cmz}; the methods for
building halo catalogues and merger trees are described in \cref{sec:hbt} and
\cref{sec:mt}, respectively.  \Cref{sec:dnf} describes how we  quantify
environmental effects on screening the fifth force. Our results are presented
in \cref{sec:results}.  Halo selection is outlined in \cref{sec:filter}, and
the processing (fitting density profiles and estimating formation times) is
covered in \cref{sec:prof,sec:cmh}.  The concentration -- formation time
relation obtained from the processed simulation data is presented in
\cref{sec:rhorho}.  We explore the sensitivity of the model predictions to the
parameter choices that specify the model in \cref{sec:params}, and to the
segregation of the halo population by the effectiveness of the screening of the
gravity fifth force in \cref{sec:sep_dnf}.  Finally, in \cref{sec:conclusions},
we present our conclusions.  Results obtained by fitting \cite{einasto1965}
(rather than NFW) profiles to determine halo structural parameters are
discussed in \cref{sec:einasto}.

\section{Theory}\label{sec:theory}

%Here we give a brief theoretical background to $\fR$ gravity (\cref{sec:fr}), a description of how the halo catalogue (\cref{sec:hbt}) and merger trees (\cref{sec:mt}) are constructed from the N-body simulations, an explanation of \textcolor{red}{the $c(m,z)$ model used to predict halo concentrations (\cref{sec:cmz}), and the environmental proxies which measure the effectiveness of the screening mechanism (\cref{sec:dnf}) in $\fR$ gravity.}

\subsection{$\fR$ gravity}\label{sec:fr}

As mentioned in the Introduction, the motivation behind the original $\fR$ model was to provide an elegant theoretical explanation for the observed accelerated expansion of the Universe \citep{buchdahl1970}.  However, in practice $\fR$ models are, by construction, fine-tuned to match the expansion history of the $\Lambda$CDM Universe, which has been tightly constrained \citep{hinshaw2013,planckcollaboration2016}.  In $\fR$ gravity, the Einstein-Hilbert action is modified by adding an extra term to the Ricci scalar $R$

\begin{equation}\label{eqn:fr}
	S=\frac{1}{16\pi G}\int\mathrm{d}^4x\sqrt{-g}\left[R+\fR\right].
\end{equation}
The $\fR$ term causes an increase in the strength of the gravitational force
compared to GR.  In order to satisfy astrophysical constraints on gravity
\citep{lombriser2014a, cataneo2015, nunes2017}, the theory contains a chameleon
screening mechanism \citep{khoury2004} which means that the GR-strength force
is recovered in dense environments.

From \cref{eqn:fr} we can derive the Poisson equation for modified gravity
\begin{equation}\label{eqn:poisson}
	\frac{1}{a^2}\vec{\nabla}^{2}\phi=\frac{16\pi G}{3}\left(\rho_{m}-\bar{\rho}_{m}\right)+\frac{1}{6}\left(R\left(f_{R}\right)-\bar{R}\right),
\end{equation}
where $f_{R}={\mathrm{d}}f/{\mathrm{d}}R$ and bars on top of variables signify
background values.  The equation remains valid for
$\vert\fR\vert\ll\vert\bar{R}\vert$ and $\vert f_{R}\vert\ll 1$, both of which
hold for the model we are investigating.  Evidently, the only difference with
respect to the Newton-Poisson equation depends solely on $f_{R}$, the
derivative of $f$ with respect to $R$.  The magnitude of $f_{R}$ relative to the classical Newtonian potential, $\phi$, splits the equation into two regimes:

\begin{enumerate}

	\item $\vert f_{R}\vert \ll \vert\phi\vert$: gravity is to a good
	      approximation described by GR, with no increased strength; these regions are called ``screened''.

	\item $\vert f_{R}\vert \geq \vert\phi\vert$: the Poisson equation is enhanced by a factor of $1/3$; in these regions screening is ineffective.

\end{enumerate}

Hence, in $\fR$ models the strength of gravity is always between $1$ and $4/3$ times the GR value.  While the particular choice of $\fR$ determines the shape of the gravitational potential in the unscreened regions, it does not affect the strength of the fifth force or the effectiveness of the screening mechanism, which is only determined by the magnitude of its derivative $\vert f_{R}\vert$.  For this reason the models are characterised by $\vert \bar{f}_{R} \vert$ with, e.g., F6 denoting $\vert \bar{f}_{R} \vert = 10^{-6}$.

Astrophysical constraints limit the choices of the present day background
value of $\vert \bar{f}_{R} \vert$.  Supernovae \citep{upadhye2013}, X-ray
\citep{terukina2014} and Solar System \citep{berry2011,lombriser2014b}
observations already rule out models with $\vert\bar{f}_{R}\vert>10^{-5}$
(F5, F4, etc.).  On the contrary, cosmologies with $\vert \bar{f}_{R} \vert
\leq10^{-7}$ show negligible differences to GR in terms of structure formation.  Here we investigate the similarities and differences between the GR and F6 ($\vert \bar{f}_{R} \vert=10^{-6}$)
simulations.

The best-studied $\fR$ model, \citet[HS]{hu2007} gravity, introduces an empirical definition of $f$:
\begin{equation}\label{eqn:hs}
	\fR = -M^{2}\frac{c_{1}\left(-R/M^2\right)^n}{c_{2}\left(-R/M^2\right)^{n}+1},
\end{equation}
where $c_{1}$ and $c_{2}$ control the screening threshold, $|f_{R0}|=c_{1}/c^{2}_{2}$, and $M=H^{2}_{0}/\Omega_{\mathrm{m}}$ is determined by the cosmology through its dependence on the Hubble constant, $H_0$, and matter density parameter, $\Omega_{\rm m}$.

As the equations describing the modifications to standard gravity are non-linear, modified gravity simulations are more demanding of computational resources than their standard gravity counterparts of the same size and resolution.  However, significant progress has been made recently in numerical techniques designed specifically for this class of theories \citep{li2012c,bose2015}. We focus our analysis on the \textsc{Liminality} simulation \citep{shi2015}, a  high-resolution, N-body simulation of HS F6 modified gravity. For comparison, a GR simulation with otherwise identical cosmology is also studied. The cosmological parameters of both runs (\cref{tbl:nbody}) have been tuned to match the \texttt{WMAP9} cosmology \citep{hinshaw2013}.

\begin{table}
	\centering
	\caption{Relevant parameters of the \textsc{Liminality} N-body simulations from \protect\citet{shi2015}.}
	\label{tbl:nbody}
	\begin{tabular}
		{lll}
		\toprule
%		Parameter             &                         & Value                                        \\
		\midrule
		$\Omega_{\mathrm{m}}$ &(matter density)         & $0.281$                                      \\
		$\Omega_{\Lambda}$ &(dark energy density)       & $0.719$                                      \\
		$\Omega_b$ &(baryon density)                    & $0.046$                                      \\
		$\sigma_{8}$ &(power spectrum amplitude)        & $0.820$                                      \\
		$n_s$ &(spectral index)                         & $0.971$                                      \\
		$h$ &($H_0/[100\, {\rm km\,s^{-1}\,Mpc^{-1}}]$) & $0.697$                                      \\
		$L$ &(box side)                                 & $64h^{-1}{\mathrm{Mpc}}$                     \\
		$m_p$ &(particle mass)                          & $1.523\times 10^8h^{-1}{\mathrm{M}_{\odot}}$ \\
		$N_p$ &(particle number)                        & $512^3$                                      \\
		$z_{\mathrm{final}}$ &(final redshift)          & $0.0$                                        \\
		$z_0$ &(initial redshift)                       & $49.0$                                       \\
		$N_{\mathrm{out}}$ & (number of outputs)        & $122$                                        \\
		\bottomrule
	\end{tabular}
\end{table}

\subsection{Mass-Concentration-Redshift relation}\label{sec:cmz}

The $c(m,z)$ model tested here, first described in \citet{ludlow2016},
uses the extended Press-Schechter (EPS) formalism to approximate
the gravitational collapse of collisionless DM haloes
\citep{bond1991,mo2010}.  In EPS, the {\em collapsed
mass history}, $m(z)$, of a dark matter halo (i.e. the sum of
progenitor masses at redshift $z$ exceeding $f\times m_{200}(z_0)$)
identified at redshift $z_0$ is given by
\begin{equation}\label{eqn:eps}
\frac{m(z)}{m_0}=\mathrm{erfc}\left(\frac{\delta_{\mathrm{sc}}(z)-\delta_{\mathrm{sc}}(z_0)}{\sqrt{2(\sigma^{2}(f\times{m_0})-\sigma^{2}(m_0))}}\right).
\end{equation}
Here $m_0=m_{200}(z_0)$ is mass at the identification redshift,
$\sigma^{2}(m)$ is the variance of the density field smoothed with
a spherical top-hat window function containing mass $m$, and
$\delta_{\mathrm{sc}}(z)\approx 1.686/D(z)$ is the redshift-dependent
spherical collapse threshold, with $D(z)$ the linear growth factor.

One difference between the EPS theory and the \citet{ludlow2016}
scheme is the definition of halo formation time: EPS defines the
formation time $z_f$ as the time at which the sum of progenitor
masses more massive than $f \times m_{200}$ first exceeds a fraction
$F \times m_{200}$, where typically $f=0.01, F=0.5$
\citep[e.g.][]{lacey1993,navarro1996}.  In \citet{ludlow2016}, $F$
is not fixed for all haloes, but instead takes on a unique value:
\begin{equation} \label{eqn:F}
	F = \frac{m_{-2}}{m_{200}}=\frac{\ln(2)-1/2}{\ln (1+c)-c/(1+c)},
\end{equation}
where the right-most equation is strictly valid for an NFW profile.
For each halo, $z_f$ therefore corresponds to the redshift at which
a fraction $m_{-2}/m_{200}$ of the halo's final mass had first
assembled into progenitors more massive than $f\times m_{200}$
(where $f=0.02$).  \citet{ludlow2016} referred to this redshift as
$z_{-2}$, to annotate its explicit dependence on the characteristic
mass, $m_{-2}$.

The CMH is scale invariant in both CDM and warm
dark matter (WDM) models, and can be used to estimate $z_{-2}$ and
the corresponding critical density, $\rho_{\mathrm{crit}}(z_{-2})$.
The $c(m,z)$ model advocated by \citet{ludlow2016} exploits the
strong, linear correlation between $\rho_{\mathrm{crit}}(z_{-2})$
and $\langle\rho_{-2}\rangle$, the mean density within $r_{-2}$.
Empirically, they found $\langle\rho_{-2}\rangle = A\times
\rho_{\mathrm{crit}}(z_{-2})$, with $A\approx 400$. Once the CMH
is known, this expression can be used to compute $\langle\rho_{-2}\rangle$,
and hence infer the halo mass profile.

The model accurately reproduces the concentrations of dark matter
haloes in both CDM and WDM cosmologies. This may appear surprising
at first as dark matter haloes in WDM simulations have been found
to display different concentrations and formation times than in CDM
\citep{maccio2013,bose2016}.  However, these changes act to preserve
the $\langle\rho_{-2}\rangle-\rho_{\mathrm{crit}}(z_{-2})$ relation
seen in CDM.

It has been shown that haloes in $\fR$ cosmologies follow NFW density profiles \citep{lombriser2014a} like their GR counterparts, but with systematically higher concentrations. Their assembly histories also differ, but only slightly \citep{shi2015}. Hence, it might be expected that the relation discovered by \citet{ludlow2016} for CDM and WDM haloes in standard gravity might hold for $\fR$ haloes only under certain conditions: (i) for small values of $\vert f_{R0}\vert$, and (ii) for all haloes {\em except} low-mass objects at low redshifts, due to screening. It is therefore plausible that the above concentration -- formation time relation will not be applicable to the full population of haloes in $\fR$ gravity, and this is the hypothesis that we test here.  This breakdown could potentially be circumvented by either re-parametrising the model or segregating haloes to reflect the influence of the fifth force, which we explore later.

\subsection{Halo identification}\label{sec:hbt}

The gravitational collapse of collisionless CDM can be approximated by the spherical collapse model
(\citealt{gunn1972}; \citealt{peebles1980}; but see \citealt{ludlow2014b}).
In this model, overdensities collapse to form dark matter haloes, which are defined as isolated regions with an average matter density larger than a threshold $\Delta_{\mathrm{vir}}\approx 178\,(\approx 200)$ times the critical density \citep[Ch. 5]{mo2010}.

Because we are primarily concerned with the GR / $\fR$ comparison, we have elected to use $r_{200}$ to define halo virial radii and $m_{200}$ for the corresponding masses.  This convention follows that of \citet{ludlow2016} and is based on the fact that, while $r_{200}$ remains well-defined and is independent of the gravity model, the virial parameters vary systematically with the strength of gravity \citep{schmidt2009}.  The virial mass and radius therefore define a sphere (centred on the particle with the minimum potential energy) that encloses a mean density equal to 200 times the critical density, $\rho_{\mathrm{crit}}\left(z\right)$, and are thus labelled with the subscript 200.
%\begin{equation}\label{eqn:rho_200}
%	\langle\rho\left(r\right)\rangle\vert_{r=r_{200}}=200\times\rho_{\mathrm{crit}}\left(z\right).
%\end{equation}

Subhaloes are locally overdense regions within haloes, and are the surviving remnants of past mergers. Haloes are initially identified using a friends-of-friends (FoF) algorithm \citep{davis1985}.  The halo catalogue is then processed using an upgraded version of \texttt{HBT} \citep[][Hierarchical Bound-Tracing algorithm]{han2012}, \texttt{HBT+} \citep{han2018}, which identifies subhaloes and builds their merger trees.

\texttt{HBT+} is a publicly available\footnote{\url{https://github.com/Kambrian/HBTplus}} merger tree code, which identifies subhaloes and follows them between simulation outputs, from the earliest snapshot at which they can be identified until the final one, building a merger tree from the catalogue on-the-fly.  A list of gravitationally bound particles is created for each halo; these are used to identify a descendant (a halo at a lower redshift, sharing subhaloes), and are passed to the successive snapshot. Each halo can have one or more progenitors (haloes at a higher redshift, sharing subhaloes).  If a halo has multiple progenitors, the most massive one is selected, and it becomes the "main" (i.e. most massive) subhalo.  Other progenitors are mapped to the subhaloes which belong to the host halo.  The host halo of a subhalo is the FoF halo containing its most bound particle.  
%The schematic merger tree shown in \cref{fig:tree} is discussed in \cref{sec:mt}.

\begin{figure}
	\includegraphics[width=\columnwidth,height=\columnwidth]{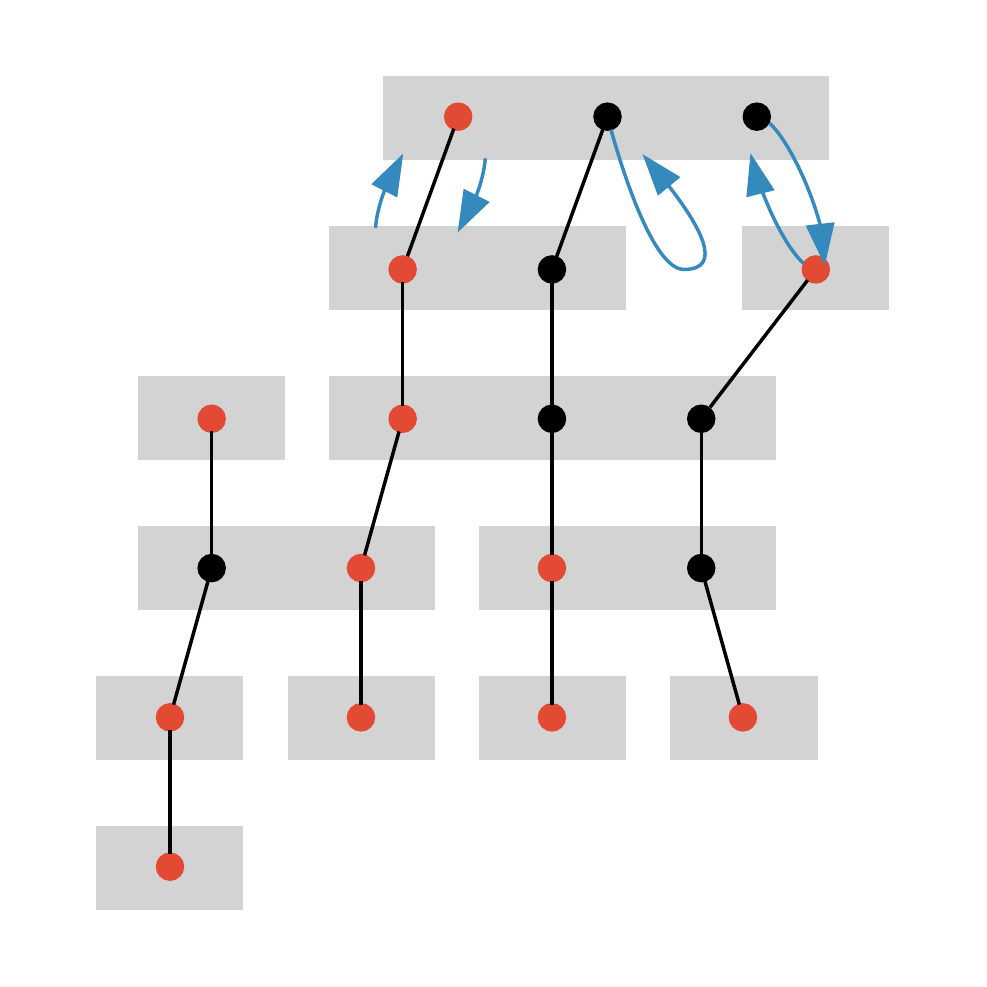}
	\caption{A schematic representation of a merger tree with two defects: a re-merger
		(halo in the second row down, on the right) and a fly-by (halo in the third
		row down on the left).  Grey rectangles represent haloes, and dots
		subhaloes; every halo has one main subhalo, marked with a red dot; subhaloes
		are matched between snapshots (black lines) by following the most bound
		particles.  The blue arrows indicate the relationships relevant in
		building merger trees, and represent (left to right): (i) halo descendat,
		(ii) halo progenitor, (iii) host of a subhalo, (iv) host of subhaloes'
		progenitor, (v) descendant of a subhalo.  This plot can be compared with
		similar diagrams included in \protect\citet{thomas2015,han2018}.
	}
	\label{fig:tree}
\end{figure}

%Simulations have shown that the density profiles of dark matter haloes can be described by simple fitting formulae, such as the NFW or \citet{einasto1965} profiles.  Here we consider an NFW density profile
%\begin{equation}\label{eqn:nfw}
%	\frac{\rho\left(r\right)}{\rho_{\mathrm{crit}}}=
%	\frac{\delta_{\mathrm{c}}}{\left(r/r_{-2}\right)\left(1+r/r_{-2}\right)^2},
%\end{equation}
%where $r_{-2}$ is the characteristic scale radius (the radius at which the
%slope of logarithmic density profile,
%$\mathrm{d}\ln\left(\rho\right)/\mathrm{d}\ln\left(r\right)\vert_{r_-2}=-2$),
%and $\delta_{\mathrm{c}}$ is a characteristic overdensity.  At fixed mass
%$m_{200}$, $\delta_{c}$ is given by
%\begin{equation}\label{eqn:delta_c}
%	\delta_{c} = \frac{200}{3}\frac{c^{3}}{\ln\left(1+c\right)-c/\l\left(1+c\right)},
%\end{equation}
%where $c$ is the dimensionless concentration parameter, defined as the ratio of $r_{200}$ over %the scale radius of a halo: $c=r_{200}/r_{-2}$.

%The other characteristic parameters of a halo, such as density ($\rho_{-2}$)
%and mass ($m_{-2}$), correspond to values calculated at radius $r=r_{-2}$.

\subsection{Merger trees}\label{sec:mt}

The merger tree of a halo, visualised in \cref{fig:tree}, can be obtained
from the \texttt{HBT+} output by following the progenitors of a given halo,
recording their host haloes, and repeating this process recursively until the
earliest progenitors are reached in each branch.  However, the trees produced by this procedure have two common defects\footnote{Technically, these are not
trees as they contain loops, and some nodes might have more than one parent
node.}:

\begin{enumerate}

	\item Re-mergers, such as the right-most halo in the second row in
	      \cref{fig:tree}, happen when one of the subhaloes temporarily becomes gravitationally
	      unbound and is identified as a separate halo for one or more snapshots; in a
	      later snapshot it merges back into the original host halo, creating a
	      ``loop''.  The halo in the ``loop" is retained as a progenitor halo and so re-mergers do not alter the collapsed mass history (which sums over
	      the masses of progenitors at any given snapshot, and as such is not
	      affected by the order or the sequence of the mergers). This is
	      similar to the scheme used to build merger trees by \citet{jiang2013}.

	\item Fly-bys (e.g. the branch merging into, and then leaving, the left-most
	      halo in the fourth row down in \cref{fig:tree}) happen when a subhalo is
	      identified as a part of a FoF halo for one or more snapshots due to a temporary spatial
	      overlap, but later becomes an isolated halo again. The presence of fly-bys pollutes the CMH, artificially inflating the mass at snapshots with extra subhaloes.

\end{enumerate}

Both defects can be avoided by simply following the main progenitor through simulation outputs -- the tree is then built by including only those progenitors that are the main subhaloes of the host in the preceding snapshot.

\subsection{Environmental screening}\label{sec:dnf}

As discussed in \cref{sec:fr}, the enhancement of gravity in the $\fR$ models
depends on the local gravitational potential.  The effectiveness of the
screening mechanism (not including self-screening) is directly related to the
environment in which the halo is found.  Following \citet{zhao2011} and
\citet{haas2012}, we use a conditional nearest neighbour distance, $D_{N,f}$,
as an environmental proxy to quantify the effectiveness of screening.

$D_{N,f}$ for a halo of mass $\bar{m}_{200}$ is defined as the distance $d$ (normalized to $\bar{r}_{200}$) to
its $N^\mathrm{th}$ nearest neighbouring whose mass, $m_{200}$, is equal to
or larger than $f \times \bar{m}_{200}$. If $D_{N,f}$ cannot be calculated (for instance, for the largest halo in a snapshot) it is assumed to be equal to
$\infty$. 

Other environment proxies, such as ``experienced gravity'' $\Phi_*$
\citep{li2011} and local spherical or shell overdensity \citep{shi2017} have
also been proposed as methods of assessing environmental impact on formation
histories. Here only $D_{N,f}$ is used since it correlates strongly with other proxies, which predict similar
local enhancements to the gravitational potential \citep{shi2017}.

\section{Results}\label{sec:results}

Our goal is to determine the relation between halo concentration
(or more specifically $\langle\rho_{-2}\rangle$) and the
critical density at the formation time $z_{-2}$ (namely
$\rho_{\mathrm{crit}}(z_{-2})$) for haloes of different masses at different
redshifts.  This section outlines the details of each step of our analysis.
The source code used for the analysis is publicly
available\footnote{\url{https://doi.org/10.5281/zenodo.2593623}}.

\subsection{Filtering \& binning}\label{sec:filter}

Our halo catalogues are obtained by filtering the \texttt{HBT+} output and retaining
objects with a minimum of 20 particles.  Since we are interested in resolving
the merger history of haloes down to progenitors with $f=0.02$ times their final mass, this places a lower limit of $n_{200}=10^3$ on the number of particles a halo must contain in order to be included in our analysis. 

Haloes are divided into bins that are equally-spaced in
$\log_{10}(m_{200}/[h^{-1}M_\odot])$, with
$\Delta \log_{10}(m_{200}/[h^{-1}M_\odot])=0.162$.  To identify potentially
unrelaxed systems we use the centre-of-mass offset parameter,
\begin{equation}\label{eqn:d_off}
	d_{\mathrm{off}}=\frac{\lvert \mathbf{r}_{p}-\mathbf{r}_{\mathrm{CM}}\rvert}{r_{200}},
\end{equation}
where $\mathbf{r}_{p}$ is the centre of potential, and $\mathbf{r}_{\mathrm{CM}}$ the centre-of-mass \citep{thomas2001,maccio2007,neto2007}.  Only haloes with
$d_{\mathrm{off}}<0.07$ are retained for analysis.

The fitting of mass profiles (\cref{sec:prof}) and calculation of formation times (\cref{sec:cmh}) is performed on the median mass profiles and CMHs, respectively, for each mass bin.

\subsection{Fitting mass profiles}\label{sec:prof}

\begin{figure}
	\includegraphics[width=\columnwidth,height=\columnwidth]{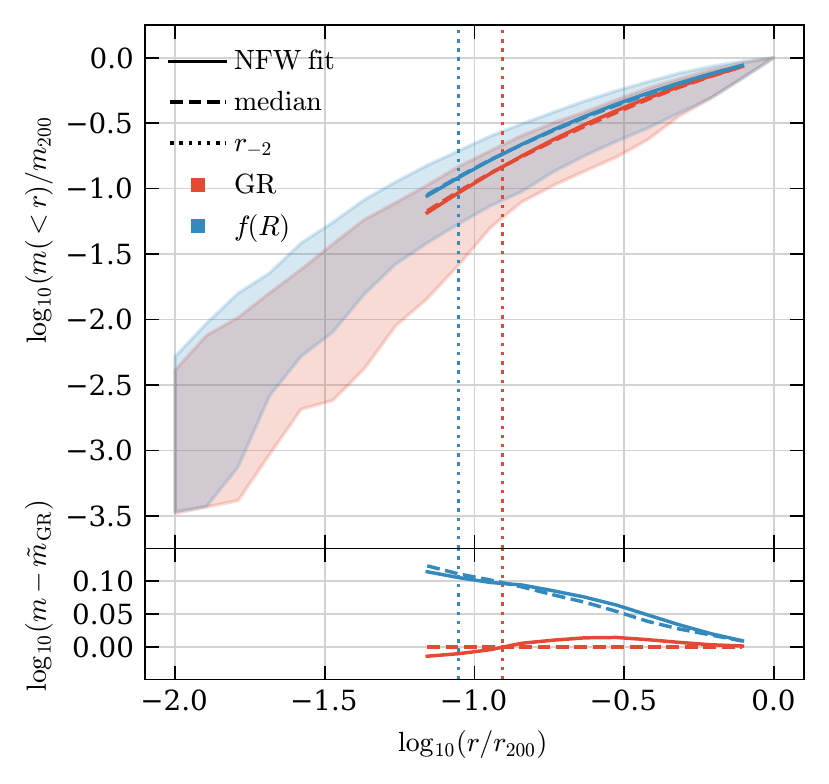}
	\caption{Radial enclosed mass profiles for haloes in the mass range
	$11.5<\log_{10}\left(m_{200}/[h^{-1}M_{\odot}]\right)<11.7$ at $z_{0}=0$.
	GR and $\fR$ runs are shown using red and blue curves, respectively, as
	indicated in the legend; residuals from GR are shown in the lower panel.
	The faint shading shows the envelope of the individual mass profiles; dashe
	lines show median mass profiles; solid lines show the best fitting NFW
	profiles to the median mass profiles, for radii between
	$r_\mathrm{min}<r<r_\mathrm{max}$; vertical dotted lines show the
	characteristic scale $r_{-2}$.  Residuals are taken from the median mass
	profile of GR haloes, $\tilde{m}_\mathrm{GR}$.
	}
	\label{fig:prof}
\end{figure}

The cumulative mass profile is defined using all particles within $r_{200}$, and not only those deemed {\em bound} to the main halo or its subhalos. These particles are assigned to logarithmically spaced radial bins, within which enclosed masses are computed. The mass profiles of all haloes found in this way are then stacked in each mass bin and the median is calculated.  Finally, the median mass profile is normalised by the total median enclosed mass, $m_{200}=m\left(r<r_{200}\right)$.  The best-fitting value of the
concentration, $c$, is obtained by minimising
\begin{equation}\label{eqn:chi}
	\chi^2=\sum_{i=0}^{20}\left[\log_{10}\left(m_{i}\right)-\log_{10}\left(m\left(r<r_{i},c\right)\right)\right]^2,
\end{equation}
where $m_i$ is the mass measured within $r_i$, $m\left(<r,c\right)$ is the mass enclosed within radius $r$ for an NFW profile with a concentration $c$ (\cref{eqn:nfw}); quantities with subscript $i$ refer to the $i^{\rm th}$ bin in $\log_{10}$ radius from the halo centre.

We have used both NFW and Einasto profiles in our analysis. Results for NFW profiles are provided in the main text and Einasto profiles are discussed in \cref{sec:einasto}, for completeness. \cref{sec:einasto} shows that the quality of fit does not improve sufficiently to warrant using the Einasto profile (which has an extra parameter) over NFW. We emphasise that the choice of analytic density profile does not change our results or conclusions.

Our fits to \cref{eqn:chi} are minimised over the radial range $r_{\min} < r_{i} < r_{\max}$, where  $r_{\min}$ is a minimum fit radius, and $r_{\max}$ is set to $0.8\times r_{200}$ to
exclude the unrelaxed outer edges of haloes \citep{ludlow2010}. We consider two definitions of $r_{\rm min}$:

\begin{enumerate}
    \item half of the mean particle separation within      $r_{200}$ \citep{moore1998},
	      \begin{equation}\label{eqn:r_min}
		      r_{\min}=\frac{1}{2}\l\left(\frac{4\pi}{3n_{200}}\right)^{1/3}r_{200},
	      \end{equation}
	      where $n_{200}$ is the number of particles enclosed within $r_{200}$, and

	\item the radius at which the two-body relaxation time is equal to the age of the universe, $t_0$ \citep{power2003,ludlow2018}, which can be approximated by the solution to
	      \begin{equation}
		      \frac{t_\mathrm{relax}\left(r\right)}{t_0}=\frac{\sqrt{200}}{8}\frac{n\left(<r\right)}{\ln\left(n\left(<r\right)\right)}\left(\frac{\langle\rho\left(<r\right)\rangle}{\rho_{\mathrm{crit}}}\right)^{-1/2}.
	      \end{equation}
		  Here $n\left(<r\right)$ is the number of particles enclosed by radius $r$
		  and $\langle\rho\left(<r\right)\rangle$ is the mean enclosed density,
		  $\langle\rho\left(<r\right)\rangle=3m\left(<r\right)/4\pi r^3$.

\end{enumerate}

Although we have considered both options, results are shown for the \citet{moore1998} definition as it is typically more conservative than the alternative. Henceforth, all $r_{\min}$ values are calculated using \cref{eqn:r_min}.

Once $c$ is found, $m_{-2}$ can be calculated from \cref{eqn:F}; the characteristic density of the halo is then given by $\langle\rho_{-2}\rangle=3\,m_{-2}/4\,\pi\,r_{-2}^3$.

\subsection{Calculating halo formation times}\label{sec:cmh}

\begin{figure}
	\includegraphics[width=\columnwidth,height=\columnwidth]{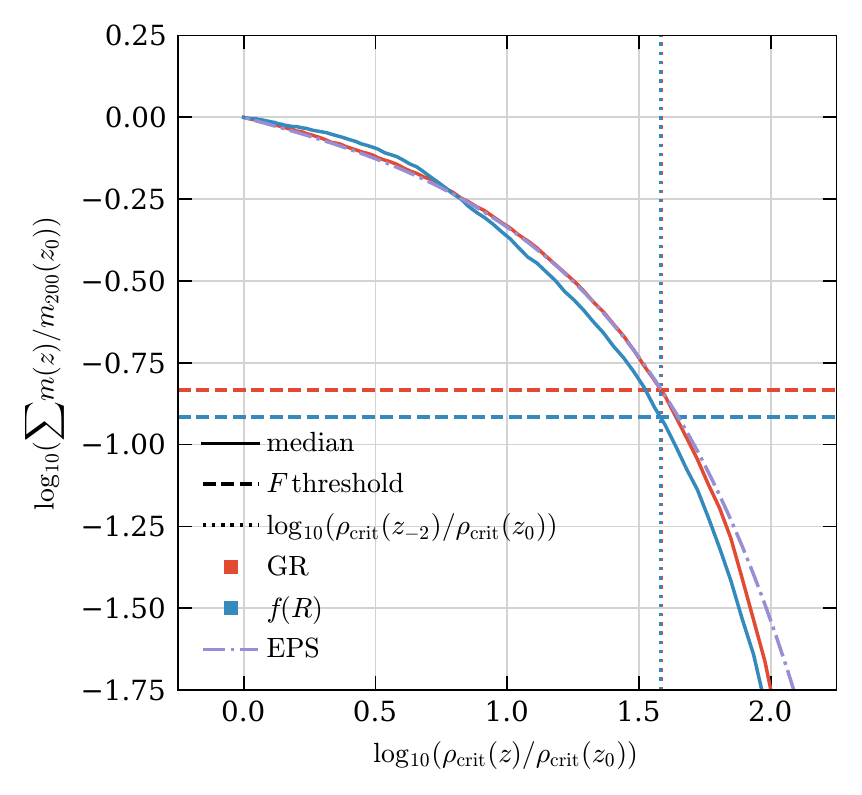}
	\caption{Median collapsed mass histories
	(CMHs) for haloes in the mass range
	$11.5<\log_{10}\left(m_{200}/[h^{-1}M_{\odot}]\right)<11.7$ at
	$z_{0}=0$.  As with \cref{fig:prof}, GR and $\fR$ runs are
	shown using red and blue lines, respectively. Solid lines
	show the median collapsed mass histories; dotted vertical
	lines indicate the formation times, $z_{-2}$, at which the
	CMHs drop below a fraction $F=m_{-2}/m_{200}$ of the virial
	mass at $z_{0}$ (shown using horizontal dashed lines).  EPS
	The purple dashed-dot line shows the EPS prediction from
	\cref{eqn:eps} for $m_0$ equal to median mass in this bin.
	}
	\label{fig:cmh}
\end{figure}

The mass growth history of a dark matter halo, $m\left(z\right)$, can be
defined in different ways. The mass assembly history (MAH) is the mass history
of a halo obtained using a ``greedy'' algorithm, by following the most massive
(or main) progenitor through all snapshots and storing its $m_{200}$. As
discussed  previously, the collapsed mass history (CMH) is defined as the sum
of the masses $m_{200,i}$ of every progenitor $i$ whose virial mass exceeds
$f\times m_{0}$,
%\begin{equation}\label{eqn:f}
%$m_{200,i} > f \times m_0$,
%\end{equation}
where $f$ is a model parameter (we use $0.02$ as our default value, but
consider alternatives as well), and $m_{0}$ is $m_{200}$ of the root halo. The
CMH therefore takes into account all branches of the merger tree at a given
snapshot.

The CMH can be obtained by querying the merger tree recursively, grouping all
progenitors and summing over the grouped masses.  However, for performance
reasons, in practice the step of building a tree can be skipped in favour of
searching for all progenitors of a root halo at each preceding snapshot. In
other words, since the halo masses are summed over, it is not the structure of
the merger tree that matters but its members.

Once the median CMH is calculated for each mass bin, it is normalised
by the final mass $m_0$ at redshift $z_0$.  For each mass bin, a
formation time $z_{-2}$ can then be calculated.  This is defined
as the time at which the CMH first exceeds a fraction $F=m_{-2}/m_0$
of the final mass, $m_0$ (\cref{eqn:F}):
\begin{equation}\label{eqn:z_form}
z_{-2}=z\ni\frac{m\left(z\right)}{m\left(z_{0}\right)}=F.
\end{equation}

The formation time may be ill-defined for non-monotonic assembly histories. The
monotonic behaviour of the CMH, while difficult to guarantee for individual
haloes, is in practice obtained by considering the median values for many
haloes in logarithmically spaced mass bins. As simulations have a finite number
of outputs, and hence finite time resolution, the value of the formation time
is obtained using linear interpolation between the snapshots which are
immediately before and after the crossing of the formation threshold fraction.

% Following \citet{navarro1997} and \citet{ludlow2016}, we focus on
% the CMH only, and hence every time a ``formation time'' is mentioned,
% this refers to $z_{\mathrm{form}}$ calculated from the CMH.

Examples of the median CMHs for $z=0$ haloes in a narrow bin of
$m_{200}$ are shown in \cref{fig:cmh}.  Solid red curves correspond
to our GR simulation, and blue to $\fR$.  An analytic
prediction from \cref{eqn:eps}, as discussed in \citet{ludlow2016},
is plotted in a purple dashed-dot line; the result agrees quite
well with the CMHs obtained from \textit{both} simulations. For
example, the formation times, $z_{-2}$ (vertical dotted lines of
corresponding color), agree with one another to $\approx 5\%$.
Nevertheless, despite similarities in CMHs, these haloes \textit{do
not} have similar concentrations. The horizontal dashed lines
correspond to $m_{-2}/m_{200}$, which show clear differences; indeed,
concentration is $30\%$ larger in $\fR$ than in GR.

%While th.is relation holds for the haloes in the GR simulation, producing a linear relation as discussed in \cref{sec:rhorho}, in F6 this prediction cannot be robustly made at all redshifts and or all masses, as illustrated in \cref{fig:cmh}.

\subsection{The density--density relation}\label{sec:rhorho}

\begin{figure}
	\includegraphics[width=\columnwidth,height=\columnwidth]{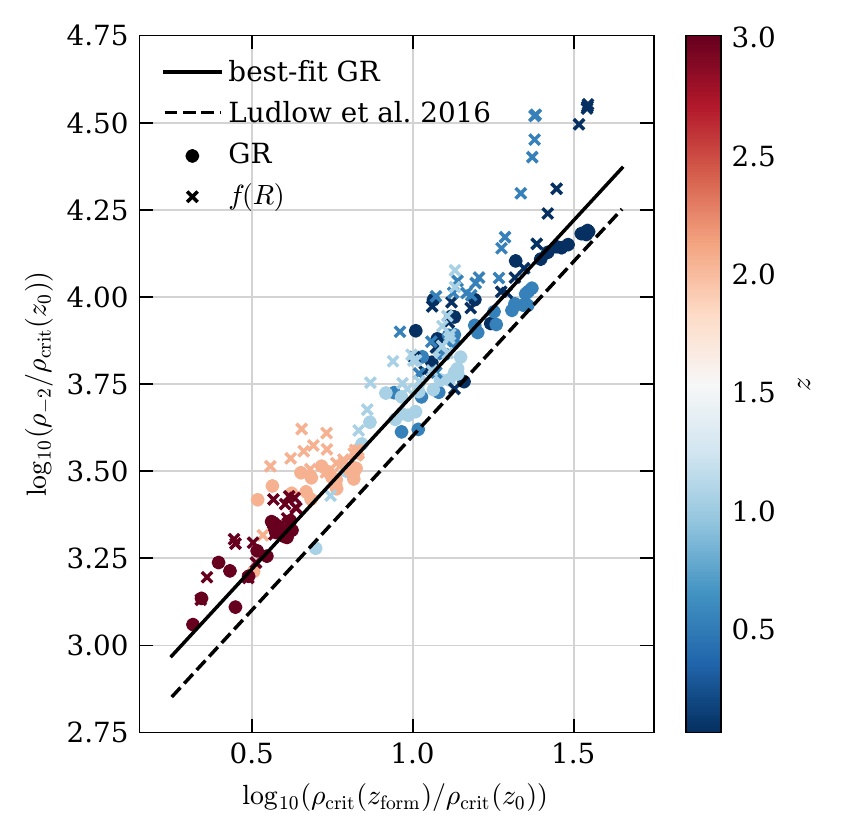}
	\caption{Mean enclosed density $\langle\rho_{-2}\rangle$ within the characteristic radius, $r_{-2}$, versus the critical density at the formation redshift, $\rho_{\rm crit}(z_{-2})$, at which a fraction $F=m_{-2}/m_{0}$ of the root halo mass $m_0$ was first contained in progenitors more massive than $f \times m_0$. Each point corresponds to median value in a logarithmically-spaced mass bin at the identification redshift $z_0$.  All densities are normalised by $\rho_{\mathrm{crit}}\left(z_0\right)$, the critical density at $z_0$.  Point types indicate the results from different gravities, as labelled. Colours indicate the identification redshift, as shown by the colour bar.  Also plotted are two lines: a dashed black one which shows the \protect\citet{ludlow2016} scaling relation $\langle\rho_{-2}\rangle=400\times\rho_{\mathrm{crit}}(z_{-2})$, and a solid black one for the best-fitting GR relation $\langle\rho_{-2}\rangle=525\times\l\rho_{\mathrm{crit}}(z_{-2})$.
	}
	\label{fig:rhorho_z}
\end{figure}

\begin{figure}
	\includegraphics[width=\columnwidth,height=\columnwidth]{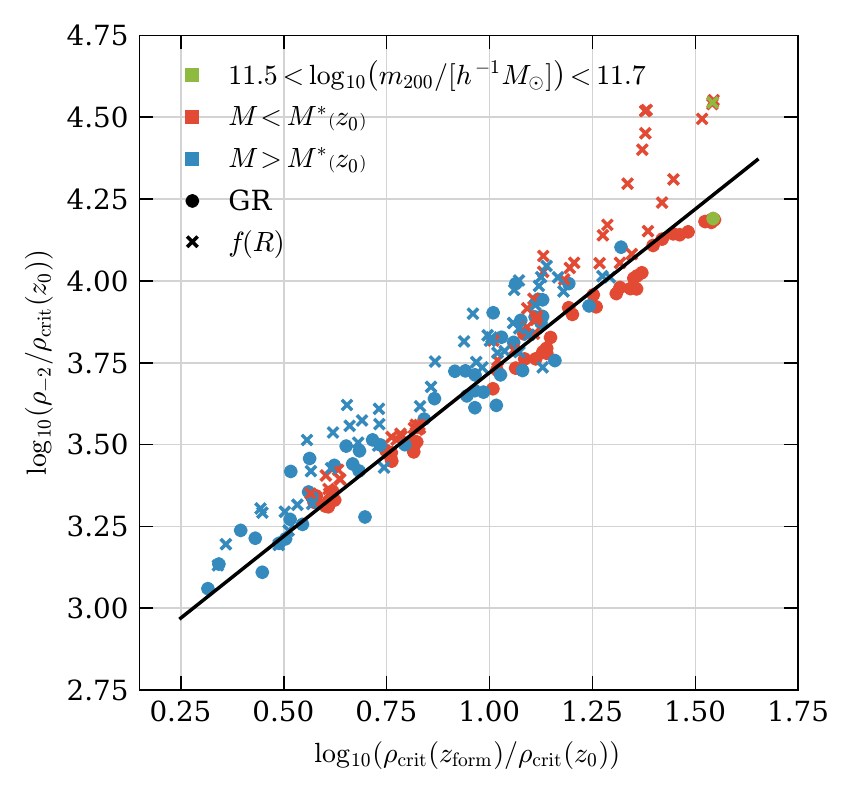}
	\caption{Same as \protect\cref{fig:rhorho_z}, but colour--coded to
	indicate different halo mass ranges. The halo population has been split into two samples: one 
	above and ones below the characteristic mass, $M^{*}\left(z_{0}\right)$,
	defined as
	$\delta_{\mathrm{sc}}\left(z_{0}\right)/\sigma\left(M^{*}\left(z_{0}\right)\right)=1$
	\protect\citep[Eq. 7.48]{mo2010}.  The mass bin containing haloes from
	\protect\cref{fig:prof,fig:cmh} at $z_{0}$ is highlighted in green.}
	\label{fig:rhorho_mstar}
\end{figure}

The above analysis was carried out at $z_{0}=0$, 0.5, 1, 2 and 3.  At each
snapshot, haloes were filtered as described in \cref{sec:filter}, and binned
into 20 logarithmically spaced mass bins spanning the range
$\log_{10}(m_0/[h^{-1}M_\odot])=11.18\;\mathrm{to}\;14.42$.  Median mass
profiles and CMHs of haloes, normalised by $m_{0}$, were used to calculate the
concentration, $c$, and formation time, $z_{-2}$, for each $m_0$ and $z_0$.
These were then converted to their equivalent values in ``density space'': $c$
expressed in terms of the characteristic density $\langle\rho_{-2}\rangle$
(following \cref{eqn:nfw}), and $z_{-2}$ in terms of the critical density,
$\rho_{\mathrm{crit}}\left(z_{-2}\right)$; both are then normalised by
$\rho_{\mathrm{crit}}\left(z_{0}\right)$.

%, set by the background cosmology as
%\begin{equation}\label{eqn:rho_c}
%	\rho_{\mathrm{crit}}\left(z\right)=\frac{3H^{2}\left(z\right)}{8\pi G}.
%\end{equation}

As shown in \cref{fig:rhorho_z,fig:rhorho_mstar}, the
$\langle\rho_{-2}\rangle-\rho_{\rm crit}(z_{-2})$ relation for F6 haloes is
similar to that in GR for most densities, but displays a steepening at high
formation redshifts where $\langle\rho_{-2}\rangle$ increases more rapidly than
$\rho_{\mathrm{crit}} (z_{-2})$. This effect is most apparent at lower
redshifts (\cref{fig:rhorho_z}) and for lower masses (\cref{fig:rhorho_mstar}).
For instance, only $\fR$ halo mass bins with
$\log_{10}(m_{200}/[h^{-1}M_\odot])\lesssim{11.9}$ at $z_0=0.5$, and with
$\log_{10}(m_{200}/[h^{-1}M_\odot])\lesssim{12.2}$ at $z_0=0$ have
$\log_{10}\left(\rho_{-2}/\rho_{\mathrm{crit}}\left(z_0\right)\right)>4.25$, as
shown by \cref{fig:rhorho_z,fig:rhorho_mstar}.  This is consistent with the
results found by \citet{shi2015} for the concentration-mass and formation
time-mass relations: while the formation times show small systematic
differences between GR and F6, the biggest discrepancy between the two is in
the form of the concentration-mass relation at low halo masses.

The concentrations recovered in the F6 model are \textit{higher} for lower mass
haloes than in GR, as demonstrated by \cref{fig:prof}; this change is in the
opposite sense to that seen on changing CDM for WDM. In both WDM and F6,
however, low mass haloes systematically form later than their GR counterparts.
In F6 gravity, although there is a systematic delay in formation histories for
low-mass haloes, it is not captured by the formation time defined as in
\cref{eqn:z_form}.

It follows that, while in WDM the formation time-concentration relation is
the same as it is in CDM (when $z_{\rm form}$ is appropriately
defined), this is not the case in $\fR$ gravity.  Even a model with an
effective screening mechanism, such as F6, affects the low mass haloes
identified at late times; these objects have slightly delayed formation times
and notably higher concentrations, which leads to the differences between
F6 and GR shown in \cref{fig:rhorho_z,fig:rhorho_mstar}.

Finally, we note that the $\langle\rho_{-2}\rangle-\rho_{\rm crit}(z_{-2})$ relation found in the GR
simulation is very similar to the one reported by \citet{ludlow2016}, but with a higher intercept value of $\approx 525$, as shown by the solid line in \cref{fig:rhorho_z}. We do not pursue this difference further.

\subsection{Sensitivity to variation of model parameters}\label{sec:params}

\begin{figure*}
	\includegraphics[width=\textwidth,height=\columnwidth]{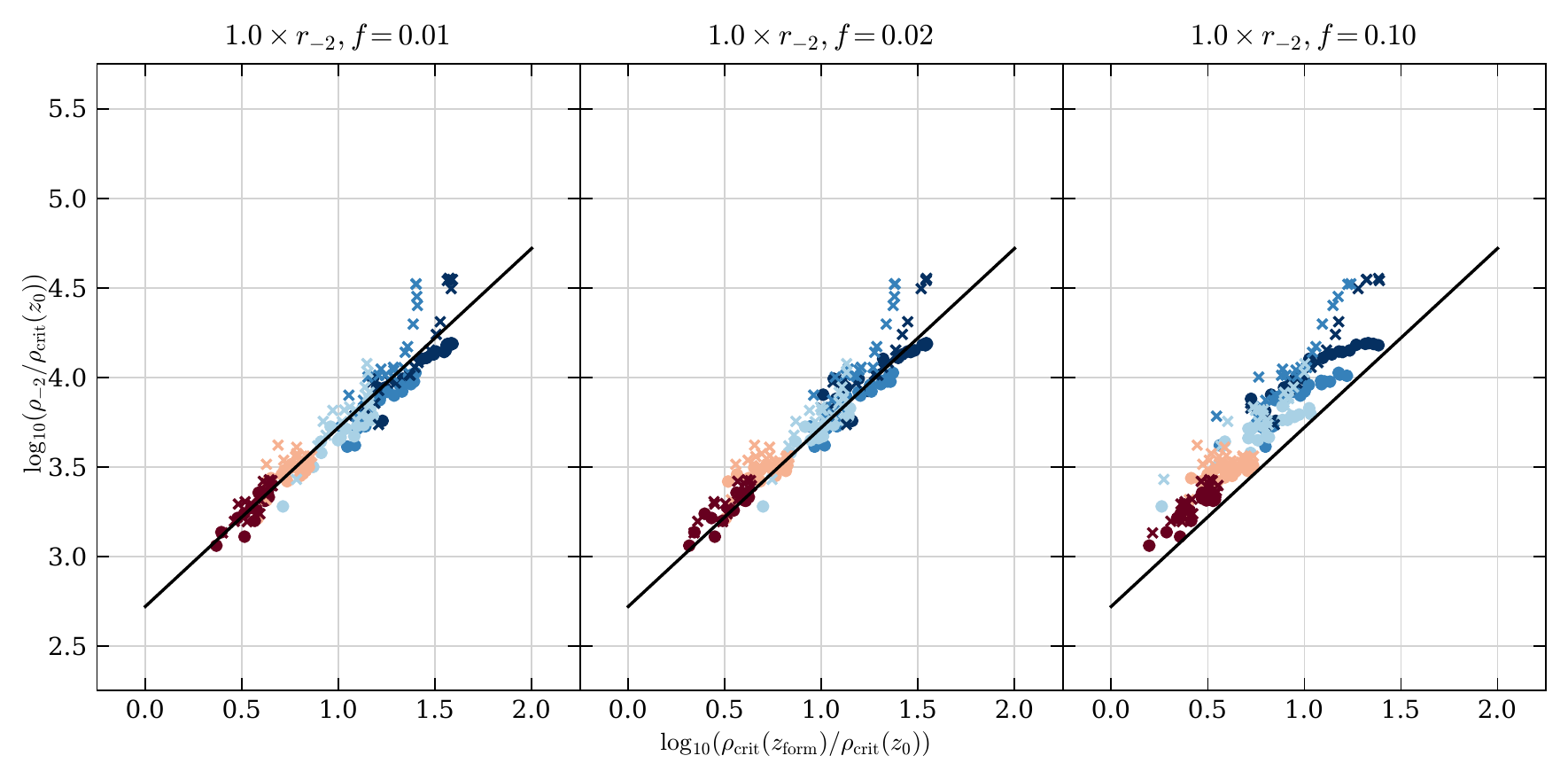}
	\caption{Like \protect\cref{fig:rhorho_z}, but with different panels
	showing different collapsed mass history parameter $f$, as labelled above
	each. The solid black line shows the best-fitting GR relation,
	$\langle\rho_{-2}\rangle=525\times\rho_{\mathrm{crit}}$, and is included for
	comparison.
	}
	\label{fig:rhorho_var_NFW_f}
\end{figure*}

\begin{figure*}
	\includegraphics[width=\textwidth,height=\columnwidth]{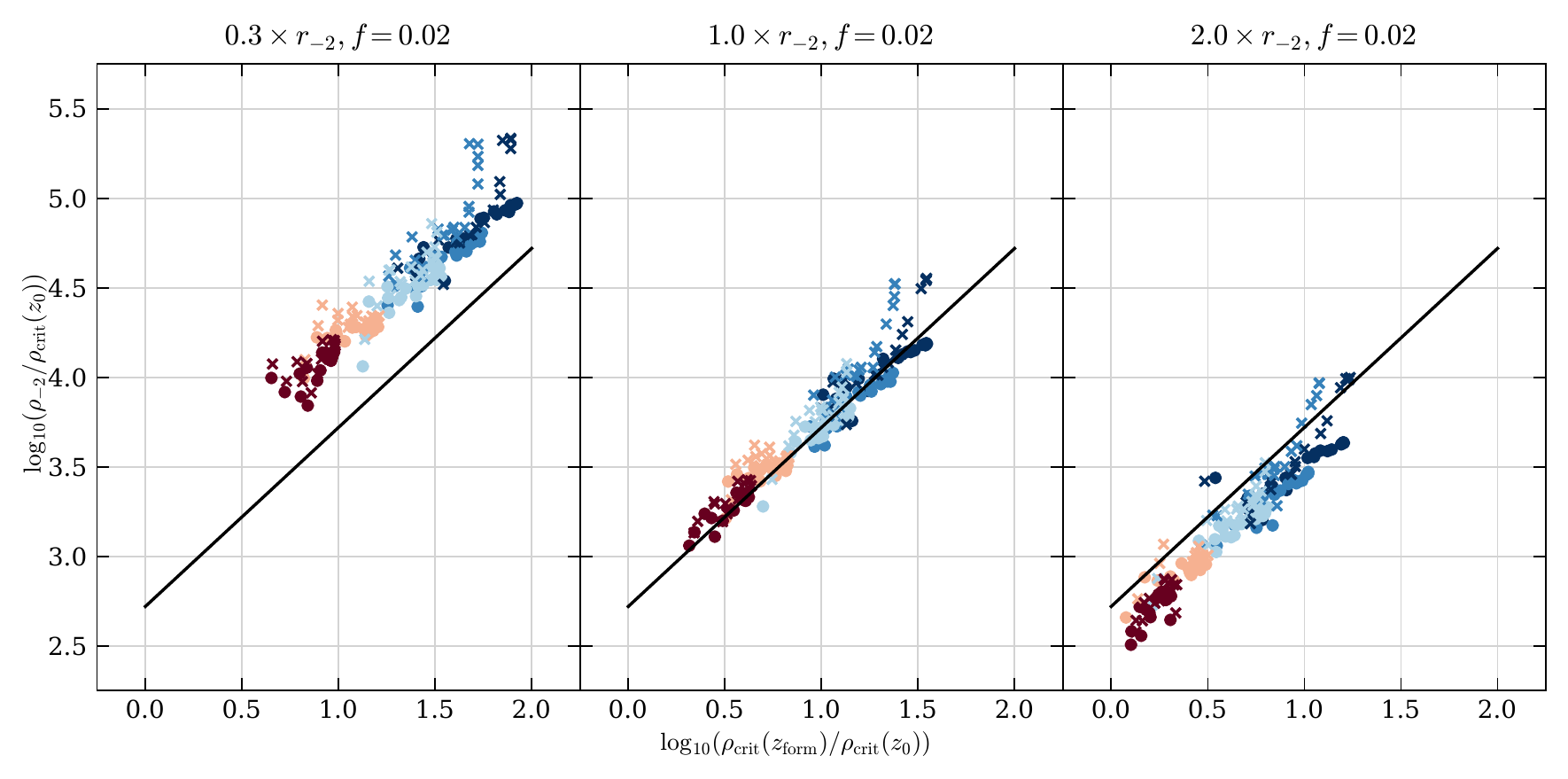}
	\caption{Like \protect\cref{fig:rhorho_z}, but with different panels
	showing mean density at different fractions of a characteristic radius
	$r_{-2}$.  The solid black line shows, for comparison, the best-fitting GR relation,
	$\langle\rho_{-2}\rangle=525\times\rho_{\mathrm{crit}}$.
	}
	\label{fig:rhorho_var_rs_f}
\end{figure*}

The parameters used to construct the CMHs (and hence to estimate $z_{-2}$) and to define halo characteristic densities can be varied to assess their impact the form of the $\langle\rho_{-2}\rangle-\rho_{\rm crit}(z_{-2})$ relation, and to potentially improve our understanding of the origin of the difference between F6 and GR.  A few such variations have been performed: first, we modify the radius defining halo characteristic densities (using $0.3\times r_{-2}$ and $2.0\times r_{-2}$), and second, the mass threshold $f$ of progenitors included in the CMH (which is varied from $0.01$ to $0.1$).

The results, presented in \cref{fig:rhorho_var_NFW_f,fig:rhorho_var_rs_f},
confirm our intuition: increasing the progenitor mass used to construct the CMHs (by increasing $f$) brings the formation time closer to the identification time, $z_0$ (the difference is more pronounced at lower redshifts, due to the normalisation used), while increasing the radius within characteristic densities are defined decreases the mean enclosed density \textit{and} brings the formation time closer to the identification redshift.  While the parameters can be tweaked to decrease the scatter and remove the time dependence of the relation (see, e.g., Figures B1 and B2 of \citealt{ludlow2016}) the $\fR$ haloes still exhibit a strong upwards trend in their concentrations--as well as a larger scatter than their GR counterparts--for all parameter combinations.  This is driven by the changes to both the $c(m,z)$ relation, and also to changes in the  mass--formation time relations, which cannot be accounted for by varying the parameters mentioned above. In $\fR$ gravity, however, the halo growth and structure are also determined by the local environment.  It is therefore important to attempt to account for local effects using an environmental proxy.

\subsection{Separation of haloes by screening}\label{sec:sep_dnf}

\begin{figure}
	\includegraphics[width=\columnwidth,height=\columnwidth]{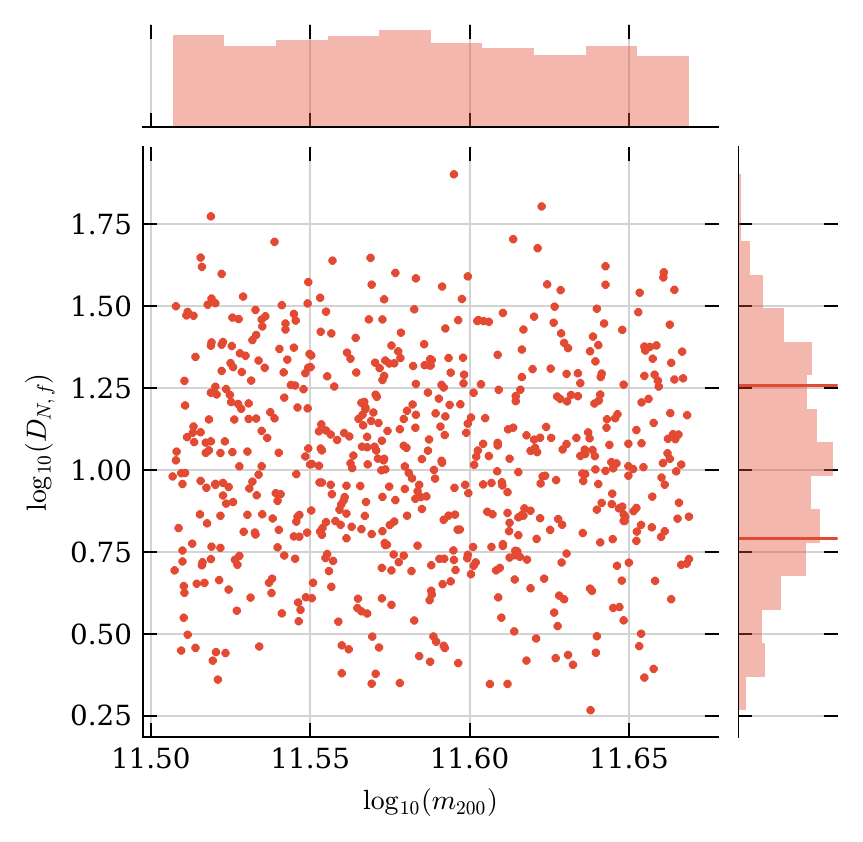}
	\caption{Environmental proxy $D_{N,f}$ ($N=1$, $f=1$) versus halo mass,
	$m_{200}$, for haloes in an example mass bin,
	$11.70<\log_{10}\left(m_{200}/[h^{-1}M_{\odot}]\right)<11.83$, at redshift $z=0$.
	Distributions of $\log_{10}\left(m_{200}/[h^{-1}M_{\odot}]\right)$ and
	$\log_{10}\left(D_{N,f}\right)$ are shown at the top- and right-hand panels,
	respectively. The two red
	lines on the $\log_{10}\left(D_{N,f}\right)$ histogram on the right
	indicate the $25^{\rm th}$ and $75^{\rm th}$ percentiles.
	}
	\label{fig:dnf}
\end{figure}

\begin{figure*}
	\includegraphics[width=\textwidth,height=\columnwidth]{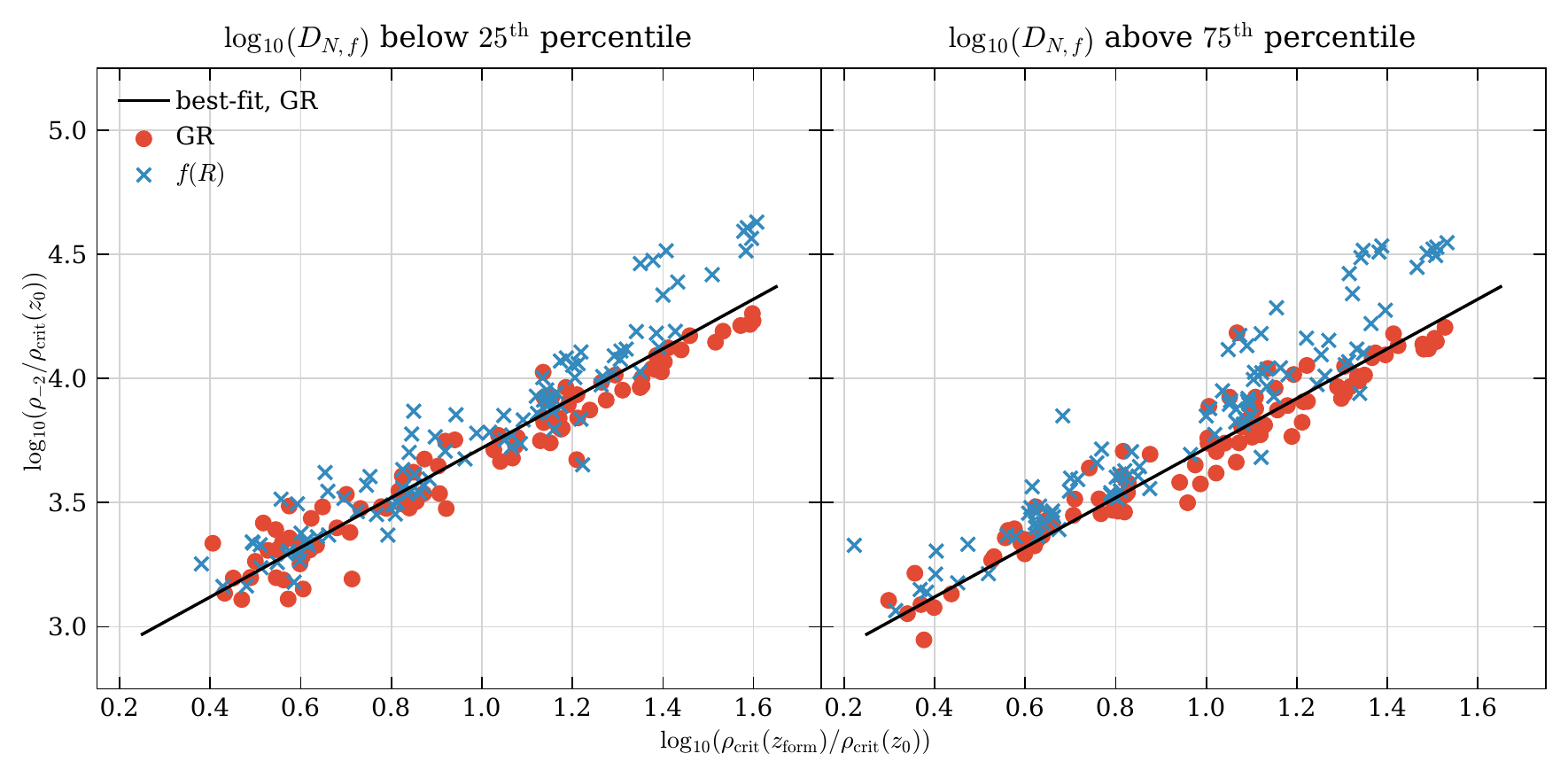}
	\caption{Like \protect\cref{fig:rhorho_z}, but split into two populations
	by the environmental proxy $D_{N,f}$.  The left panel shows the relation
	for bins including haloes below the $25^{\rm th}$ percentile; the right
	panel shows the same relation for bins including haloes above the $75^{\rm
	th}$ percentile. Colours and symbols distinguish between gravity models:
	red circles represent GR and blue crosses F6.  Both panels include
	the best fitting GR relation
	$\langle\rho_{-2}\rangle=525\times\rho_{\mathrm{crit}}$ (solid black line)
	for reference (note that the fit is performed over the full population,
	regardless of the environmental proxy).
	}
	\label{fig:rhorho_cut}
\end{figure*}

As discussed in \cref{sec:fr}, $\fR$ gravity only affects haloes which are
outside screened regions, while the screened ones grow in a manner that is
largely indistinguishable from GR.  It is clear from \cref{fig:rhorho_z} that
low mass haloes are typically the ones displaying the most prominent differences
between the two simulations, implicating the fifth force as the root cause.
However, it is natural that each mass bin contains both screened and unscreened
objects.  By using $D_{N,f}$ with $N=1$, $f=1.0$ as an environmental proxy (see
\cref{sec:dnf}), we have attempted to separate haloes inside each mass bin into
two populations, quantifying how strong the environmental screening effect
should be.

The $D_{N,f}$ values have been calculated for each halo at each redshift.  Here
we consider the distribution of $D_{N,f}$ in bins of halo mass focusing on the
extremes of the distribution which we expect will show the biggest contrast in
the efficiency of screening.  The halo population at each redshift is split
into two sub-groups: those below the $25^{\rm th}$ and above the $75^{\rm th}$
percentiles.  The most massive object, with $D_{1,1}=\infty$, is excluded.  The
$\langle\rho_{-2}\rangle-\rho_{\rm crit}(z_{-2})$ relations were then
recalculated (again stacked by mass) for the two sub-groups separately, and are
presented in \cref{fig:rhorho_cut}.

It is to be expected that the haloes with the lowest values of $D_{N,f}$, which
are the ones that are closest to objects of comparable masses and hence in the
highest density environments, will follow a concentration-formation relation
closest to that displayed by GR haloes, since they are screened from the
enhanced gravity. Haloes with high-$D_{N,f}$ may display a different power-law,
as seen in \cref{fig:rhorho_z}.  However, as clearly demonstrated in
\cref{fig:rhorho_cut}, while selecting haloes by their $D_{N,f}$ value has
little to no effect on the GR relation, it also has little impact on the F6
haloes.  This suggests that this difference cannot be easily accounted for by a
local environment proxy (which excludes effects such as self-screening and
transitioning from screened to unscreened regions) and is driven by some other aspect of the rich $\fR$ phenomenology.

\section{Conclusions}\label{sec:conclusions}

We have compared two high resolution dark matter only
simulations, one using GR and the other F6 gravity.  We constructed
collapsed mass histories of haloes using their merger trees obtained from \texttt{HBT+} \citep{han2018}.  We then binned the haloes by mass and stacked their enclosed mass profiles, $m(r)$, and CMHs to obtain median concentrations, $c$, and formation times, $z_{-2}$, which we used to construct the $\langle\rho_{-2}\rangle-\rho_{\rm crit}(z_{-2})$ relation. This relation is linear in GR--and hence may be used to predict concentrations when CMHs are known--but not in F6. The differences are primarily due to a relative {\em enhancement} of concentration for low-mass objects in F6 which have slightly {\em delayed} formation times times relative to GR.

We have made several attempts to recover a linear relation
from the results of the F6 simulation. For example, we varied
the free parameters of the model (i.e. the fraction $f$ of the final halo mass
that a progenitor must exceed to be included in the CMH, and fraction of the characteristic radius $r_{-2}$ used to define the characteristic densities) to find a region in the parameter space which
produces the most promising relation.  While there are values of parameters
which improve upon the conventional choice for GR ($f=0.02$, $1.0 \times r_s$),
there are trade-offs with regards to scatter and gradient of the line.
Furthermore, to account for the mixing of the screened and unscreened haloes in
each mass bin, we split the halo catalogue into two sub-populations using an
environmental proxy $D_{N,f}$, which also had little
effect.

Our overall conclusion is that the form of the concentration--formation time
relation is particular to the gravitational force in the adopted cosmological
model and its origin remains unknown.  The key difficulty seems to lie in the
question of why haloes with very similar formation redshifts can nevertheless have very different concentrations. One possibility is that the definition of formation time ($z_{-2}$) or assembly history (CMH)--which function well for GR models for $c(m,z)$--require amendments for $\fR$.

Since the relation is sensitive to model parameter variation, but not to
environment--based splitting, it would be interesting to further test the
relation for a dependence on self-screening.  This could be tested by splitting halo populations using a self-screening proxy, as
well as running the analysis on other cosmologies, such as F5, F4 and enhanced
(4/3 the conventional strength) gravity simulations.  We believe that looking
into the changes in the concentration -- formation relation in different
gravity regimes is a promising avenue of research into the nature and origin of
the correlation between halo concentrations and formation times.

\section*{Acknowledgements}

We thank Difu Shi for helpful discussions and for providing simulation outputs.
This work was supported by the Science and Technology facilities Council
ST/P000541/1.  PO acknowledges an STFC studentship funded by STFC grant
ST/N50404X/1. ADL acknowledges financial supported from the
Australian Research Council through their Future Fellowship scheme (project
number FT160100250).This work used the DiRAC@Durham facility managed by the
Institute for Computational Cosmology on behalf of the STFC DiRAC HPC Facility
(www.dirac.ac.uk). The equipment was funded by BEIS capital funding via STFC
capital grants ST/K00042X/1, ST/P002293/1, ST/R002371/1 and ST/S002502/1,
Durham University and STFC operations grant ST/R000832/1. DiRAC is part of the
National e-Infrastructure.

\bibliographystyle{mnras}
\bibliography{oleskiewicz2019a}

\appendix

\section{Einasto profile}\label{sec:einasto}

The Einasto density profile \citep{einasto1965} can be expressed as
\begin{equation}
	\ln\left(\frac{\rho}{\rho_{-2}}\right)=-\frac{2}{\alpha}\left[\left(\frac{r}{r_{-2}}\right)^{\alpha}-1\right],
\end{equation}
where $r_{-2}$ is a scale radius (at which where the logarithmic slope of the density profile is equal to $-2$), and $\alpha$ is a "shape" parameter. 

Fits using both NFW and Einasto density profiles have been performed for
comparison.  We have computed and compared model selection criteria, called AIC and BIC, as an objective way to determine if the additional parameter in the Einasto profile is justified in terms of improved fits to the simulation results \citep{akaike1974,schwarz1978}.  The AIC and BIC measures take into account the $\chi^2$ value of the fit and the number of free parameters. The fit with the smallest value of AIC or BIC is deemed to be the most appropriate one to use\footnote{There is a subtle difference between the AIC and BIC statistics.  BIC introduces a higher penalty for more complicated models; this is only important if the criteria give conflicting results, which is not the case here}.  The NFW and Einasto density profiles, and the corresponding values of the AIC and BIC statistics for an illustrative mass bin at $z_{0}=0$ are shown in \cref{fig:prof} and in \cref{tbl:fit}. Despite the fact that the Einasto profile produces a better fit, it also yields higher values of the information criteria, which indicates that the NFW profile is the
more justified choice.

\begin{figure}
	\includegraphics[width=\columnwidth,height=\columnwidth]{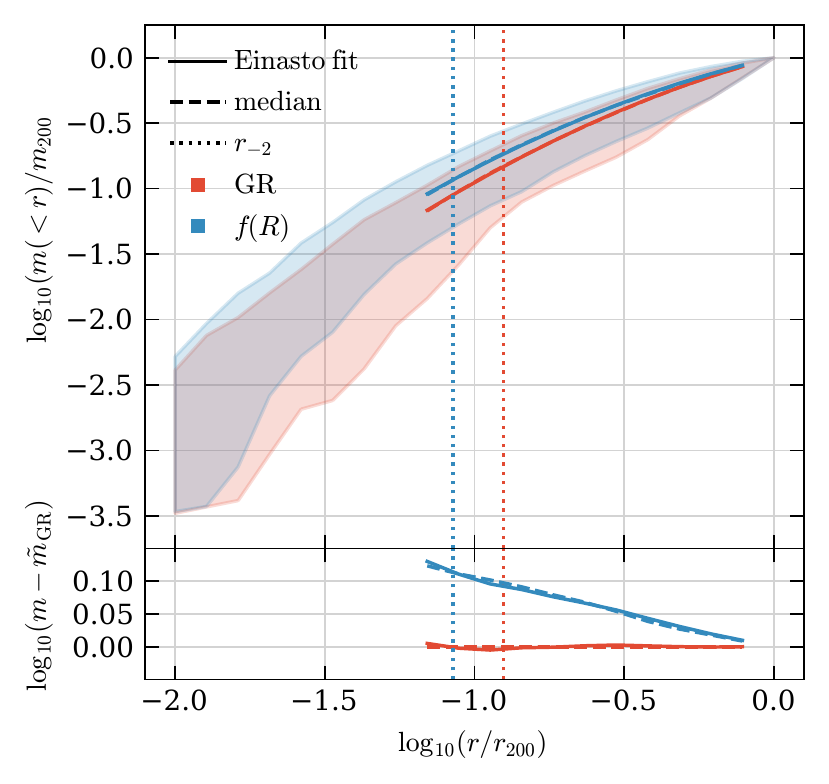}
	\caption{Like \protect\cref{fig:prof}, but for the Einasto density profile.
	}
\end{figure}

\begin{figure}
	\includegraphics[width=\columnwidth,height=\columnwidth]{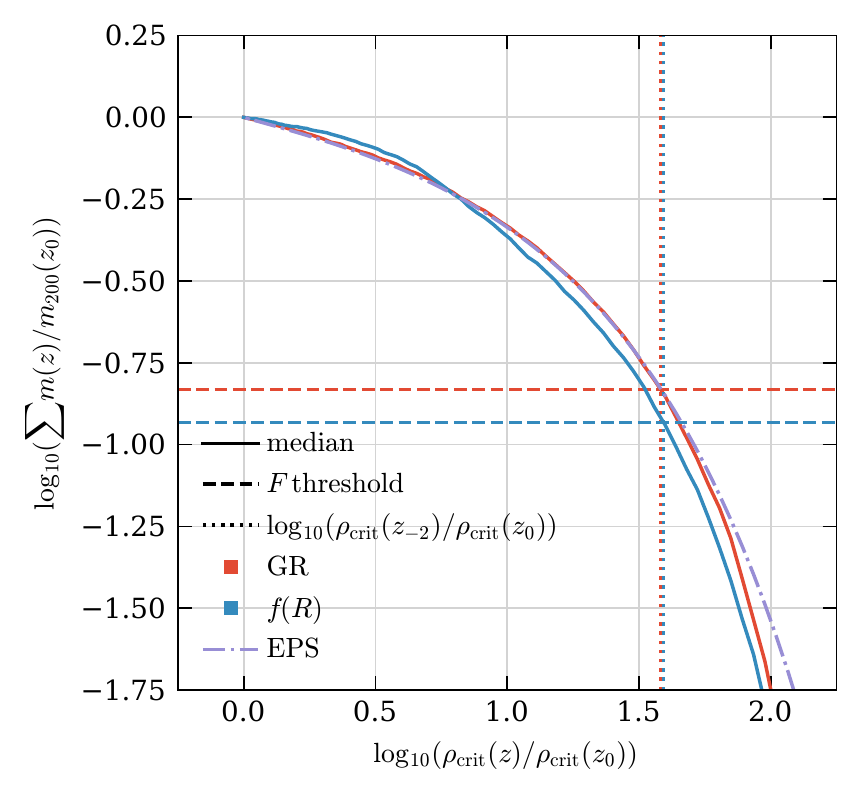}
	\caption{Like \protect\cref{fig:cmh}, but for the values of $F$ obtained using the Einasto formula.
	}
\end{figure}

\begin{table}
	\centering
	\caption{Goodness-of-fit comparison between the NFW and Einasto density
	profiles for haloes with masses in the range
	$11.5<\log_{10}\left(m_{200}/[h^{-1}M_{\odot}]\right)<11.7$ at $z_{0}=0$
	for the GR run.
	}
	\label{tbl:fit}
	\begin{tabular}
		{lll}
		\toprule
		                     & NFW   & Einasto \\
		\midrule
		number of parameters & 1     & 2       \\
		AIC                  & 2.002 & 2.399   \\
		BIC                  & 4.002 & 4.797   \\
		\bottomrule
	\end{tabular}
\end{table}

% \section{3D visualisation of environmental proxy}\label{sec:3d}

% \begin{figure*}
% 	\includegraphics[width=\textwidth,height=0.9\textwidth]{fig/3d_fr6_b64n512_122.pdf}
% 	\caption{3D visualisation of halo catalogue obtained from
% 	the $\fR$ simulation at redshift $z_0=0$, consisting of
% 	over $4000$ haloes.  The sizes of the points reflect
% 	$log_{10}\left(m_{200}\right)$, and the colours indicate
% 	the  environmental proxy $\log_{10}\left(D_{N,f}\right)$
% 	used to approximate the screening and the strength of the
% 	fifth force.  Note how the larger values of $D_{N,f}$
% 	(indicating lower environmental impact on formation \&
% 	evolution) are reserved for objects which are either more
% 	massive, or isolated \& small, while smaller values of
% 	$D_{N,f}$ are obtained for objects which are in the vicinity
% 	of the similar-mass neighbours.
% 	}
% 	\label{fig:3d}
% \end{figure*}

\bsp
\label{lastpage}
\end{document}